\documentstyle[12pt,psfig]{article}
\newcommand{\nsect}{\setcounter{equation}{0}
\def\theequation{\thesection.\arabic{equation}}\section}

\def\baselinestretch{1.5}
\catcode`\@=11
\def\marginnote#1{}
\def\ifmath#1{\relax\ifmmode #1\else $#1$\fi}

\def\bold#1{\setbox0=\hbox{$#1$}%
     \kern-.025em\copy0\kern-\wd0
     \kern.05em\copy0\kern-\wd0
     \kern-.025em\raise.0433em\box0 }

\def\GENITEM#1;#2{\par\vskip6pt \hangafter=0 \hangindent=#1
   \Textindent{$ #2$ }\ignorespaces}

\newcount\hour
\newcount\minute
\newtoks\amorpm
\hour=\time\divide\hour by60
\minute=\time{\multiply\hour by60 \global\advance\minute by-
\hour}
\edef\standardtime{{\ifnum\hour<12 \global\amorpm={am}%
    \else\global\amorpm={pm}\advance\hour by-12 \fi
    \ifnum\hour=0 \hour=12 \fi
    \number\hour:\ifnum\minute<100\fi\number\minute\the\amorpm}}
\edef\militarytime{\number\hour:\ifnum\minute<100\fi\number\minute}
\def\draftlabel#1{{\@bsphack\if@filesw {\let\thepage\relax
  \xdef\@gtempa{\write\@auxout{\string
    \newlabel{#1}{{\@currentlabel}{\thepage}}}}}\@gtempa
    \if@nobreak \ifvmode\nobreak\fi\fi\fi\@esphack}
     \gdef\@eqnlabel{#1}}
\def\@eqnlabel{}
\def\@vacuum{}
\def\draftmarginnote#1{\marginpar{\raggedright\scriptsize\tt#1}}
\def\draft{\oddsidemargin -.5truein
        \def\@oddfoot{\sl preliminary draft \hfil
        \rm\thepage\hfil\sl\today\quad\militarytime}
        \let\@evenfoot\@oddfoot \overfullrule 3pt
        \let\label=\draftlabel
        \let\marginnote=\draftmarginnote

\def\@eqnnum{(\theequation)\rlap{\kern\marginparsep\tt\@eqnlabel}%
\global\let\@eqnlabel\@vacuum}  }
\def\preprint{\twocolumn\sloppy\flushbottom\parindent 1em
        \leftmargini 2em\leftmarginv .5em\leftmarginvi .5em
        \oddsidemargin -.5in    \evensidemargin -.5in
        \columnsep 15mm \footheight 0pt
        \textwidth 250mmin      \topmargin  -.4in
        \headheight 12pt \topskip .4in
        \textheight 175mm
        \footskip 0pt

\def\@oddhead{\thepage\hfil\addtocounter{page}{1}\thepage}
        \let\@evenhead\@oddhead \def\@oddfoot{} \def\@evenfoot{}
}
\def\titlepage{\@restonecolfalse\if@twocolumn\@restonecoltrue\onecolumn
     \else \newpage \fi \thispagestyle{empty}\c@page\z@
        \def\thefootnote{\fnsymbol{footnote}} }
\def\endtitlepage{\if@restonecol\twocolumn \else  \fi
        \def\thefootnote{\arabic{footnote}}
        \setcounter{footnote}{0}}  %\c@footnote\z@ }
\catcode`@=12
\relax
\def\be{\begin{equation}}
\def\ee{\end{equation}}
\def\bea{\begin{eqnarray}}
\def\eea{\end{eqnarray}}
\def\simlt{\stackrel{<}{{}_\sim}}
\def\simgt{\stackrel{>}{{}_\sim}}

\def\NPB#1#2#3{{\it Nucl.~Phys.} {\bf{B#1}} (19#2) #3}
\def\PLB#1#2#3{{\it Phys.~Lett.} {\bf{B#1}} (19#2) #3}
\def\PRD#1#2#3{{\it Phys.~Rev.} {\bf{D#1}} (19#2) #3}
\def\PRL#1#2#3{{\it Phys.~Rev.~Lett.} {\bf{#1}} (19#2) #3}

\def\MPLA#1#2#3{{\it Mod.~Phys.~Lett.} {\bf#1} (19#2) #3}

\def\AP#1#2#3{{\it Ann.~Phys.} {\bf#1} (19#2) #3}

\def\HPA#1#2#3{{\it Helv.~Phys.~Acta} {\bf#1} (19#2) #3}
\def\JETPL#1#2#3{{\it JETP~Lett.} {\bf#1} (19#2) #3}

\def\bigint{{\displaystyle\int}}

\def\mst1{m_{\;\widetilde{t}_{1}}}

\def\st{\;\widetilde{t}}

\def\mst2{m_{\;\widetilde{t}_{2}}}
\def\mst12{m_{\;\widetilde{t}_{1,2}}}

\def\msb1{m_{\;\widetilde{b}_{1}}}
\def\msb2{m_{\;\widetilde{b}_{2}}}
\def\msb12{m_{\;\widetilde{b}_{1,2}}}

\def\mtilde2{\widetilde{m}^{2}}

\relax
\def\baselinestretch{1.25}
\begin{document}
\topmargin-2.5cm
%\draft
%\hoffset = .65 in

%\preprint
%
\begin{titlepage}
\begin{flushright}
IEM-FT-168/98 \\
hep--ph/9801272 \\
\end{flushright}
\vskip 0.3in
\begin{center}{\Large\bf Bubbles in the Supersymmetric Standard Model
\footnote{Work supported in part by the European Union
(contract CHRX/CT92-0004) and CICYT of Spain
(contract AEN95-0195).} }
\vskip .5in
{\bf J.M. Moreno,  M. Quir{\'o}s and M. Seco} \\
\vskip.35in
Instituto de Estructura de la Materia, CSIC, Serrano
123, 28006-Madrid, Spain
\end{center}
\vskip2.3cm
\begin{center}
{\bf Abstract}
\end{center}

\begin{quote}
We compute the tunneling probability from the symmetric phase to the true
vacuum, in the first order electroweak phase transition of the
MSSM, and the corresponding Higgs profiles along the bubble wall.
We use the resummed two-loop temperature-dependent
effective potential, and pay particular attention to the light
stop scenario, where the phase transition can be sufficiently
strongly first order not to wipe off any previously generated
baryon asymmetry. We compute the bubble parameters 
which are relevant for the baryogenesis mechanism: the wall thickness and
$\Delta\beta$. The two-loop corrections provide important
enhancement effects, with respect to the one-loop results, 
in the amount of baryon asymmetry.
\end{quote}
\def\baselinestretch{1.5}
\vskip3.cm

\begin{flushleft}
IEM-FT-168/98\\
January 1998 \\
\end{flushleft}
\end{titlepage}
\setcounter{footnote}{0}
\setcounter{page}{0}
\newpage

\nsect{Introduction}
Electroweak baryogenesis~\cite{baryogenesis} is an appealing mechanism to 
explain the observed 
value of the baryon-to-entropy ratio, $n_B/s\sim 10^{-10}$, at the
electroweak phase transition~\cite{first,reviews}, that can be tested at 
present and future 
high-energy colliders. Although the Standard Model contains all the
necessary ingredients~\cite{baryogenesis} for a successful baryogenesis, 
it fails in providing enough baryon asymmetry. In particular it has been
proven by perturbative~\cite{AndH,improvement,twoloop} and
non-perturbative~\cite{nonpert} methods that, for Higgs masses allowed by
present LEP bounds, the phase transition is too weakly first order, 
and any previously generated baryon
asymmetry would be washed out after the phase transition. On the other hand
the amount of CP violation arising from the CKM phase is too small for
generating the observed baryon asymmetry~\cite{CPSM}. Therefore electroweak
baryogenesis requires physics beyond the Standard Model at the weak
scale.

Among the possible extensions of the Standard Model at the weak scale, its
minimal supersymmetric extension (MSSM) is the best motivated one. It
provides a technical solution to the hierarchy problem and has its roots 
in more fundamental theories unifying gravity with the rest of interactions.
The MSSM has new violating phases~\cite{CPMSSM} that can drive enough
amount of baryon asymmetry~\cite{CQRVW}-\cite{last} provided that 
the previous phases are not much
less than 1 and the charginos and neutralinos are not heavier than 
200 GeV~\footnote{The first and second generation squarks are required
to have masses $\cal O$(few) TeV because of the experimental bounds on  
the neutron electric dilope moment.}. As for
the strength of the phase transition~\cite{early}-\cite{mariano2}, a region
in the space of supersymmetric parameters has been 
found~\cite{CQW}-\cite{CQW2} where the phase transition is strong enough to let
sphaleron interactions go out of equilibrium after the phase transition and
not erase the generated baryon asymmetry. This region (the so-called 
light stop scenario) provides values of the
lightest Higgs and stop eigenstates which are: $75$ GeV $\simlt 
m_h \simlt 105$ GeV,
100 GeV$\simlt m_{\st}\simlt m_t$. It will be covered at LEP2 and
Tevatron colliders.

In all calculations of the baryon asymmetry the details of the wall
parameters play a prominent role in the final result. In particular the wall
thickness, $L_\omega$, and the relative variation of the two Higgs fields
along the wall, $\Delta\beta$, are typical parameters which 
the generated baryon asymmetry depends upon. Although reasonable assumptions
about the Higgs profiles along the wall have been done, as e.g. kinks or
sinusoidal patterns interpolating between the broken and the symmetric phases,
as well as estimates on the value of $\Delta\beta$ based on purely potential
energy considerations, it is clear that the reliability of those estimates
as well as more precise computations of the baryon asymmetry should rely on
realistic calculations of the Higgs profiles and the tunneling processes from
the false to the true vacuum. Such a task, having been achieved in the case
of one Higgs field in the Standard Model~\cite{Brihaye93}, 
was still missing in the case of two-Higgs field models. 

In this paper we compute the tunneling probability
from the false to the true vacuum in the first order electroweak 
phase transition of the MSSM, and the corresponding Higgs profiles 
along the wall. In particular we will concentrate in the
region of the supersymmetric parameters corresponding to the light stop
scenario, where the phase transition is strong enough not to wash out the
generated baryon asymmetry. We will use the MSSM effective potential at
finite temperature including the most important two-loop corrections.
The two-loop corrections have been proven to be very important in determining
the strength of the phase transition and we will demonstrate that they
also play
a very relevant role in the baryogenesis mechanism. In particular we will
compare the results using the one-loop effective potential, that has been
used so far 
for the determination of the baryon asymmetry, with the corresponding
two-loop results. We will see that there is an important enhancement coming
from two-loop effects. The plan of the paper is as follows. In section 2
the method to compute the bubbles in the MSSM will be described in detail.
The results of our numerical analysis will be presented in
section 3. 
The possibility of bubbles involving squark fields will be explored in
section 4 and section 5 is devoted to draw our conclusions.

%\vspace{1.5cm}
\nsect{Higgs Bubbles}

The Higgs sector of the MSSM requires two Higgs doublets, with
opposite hypercharges, as
\begin{equation}
\label{higgsmssm}
H_1  =  \left(
\begin{array}{c}
H_1^0 \\
H_1^-
\end{array}
\right)_{-1/2}\ ; \qquad
H_2  =  \left(
\begin{array}{c}
H_2^+ \\
H_2^0
\end{array}
\right)_{1/2}
\end{equation}
and tree-level potential:
\begin{eqnarray}
\label{potmssm}
V^{(0)}& = & m_1^2 H_1^\dagger H_1+m_2^2 H_2^\dagger H_2
+m_3^2(H_1 H_2+h.c.) \\
&&+\frac{1}{8}g^2\left(H_2^\dagger \vec{\sigma} H_2+
H_1^\dagger\vec{\sigma}H_1\right)^2
+\frac{1}{8}g'^2\left(H_2^\dagger H_2-H_1^\dagger H_1\right)^2
\nonumber
\end{eqnarray}

The field configuration describing the tunneling in a theory 
with just one scalar field in three or more dimensions
is known to be spherically symmetric~\cite{Col78}. Here we will assume that
the bubbles driving the electroweak phase transition in the MSSM and
involving the two neutral higsses also have spherical symmetry. They
correspond to the ansatz:
\be
\label{background}
H_1(\vec{x})=h_1(r)\
\left[
\begin{array}{c}
1 \\
0
\end{array}
\right] \ ;\qquad
H_2(\vec{x})=h_2(r)\ 
\left[
\begin{array}{c}
0 \\
1
\end{array}
\right]\ , 
\ee
where $r\equiv\sqrt{\vec{x}^{\,2}}$.
In the presence of the background (\ref{background}) the
tree-level potential reads as
\be
\label{treefin}
V^{(0)}(h_1,h_2) =  m_1^2\; h_1^2(r)+
m_2^2\; h_2^2(r)+2\;m_3^2 h_1(r)h_2(r) 
+\frac{g^2+g'^2}{8}  \left[h_1^2(r)-h_2^2(r)\right]^2
\ee
and the one-loop corrections, in the 't Hooft-Landau gauge and
in the $\overline{\rm MS}$ renormalization scheme, are given by
\be
\label{oneloop}
V^{(1)}(h_1,h_2)=\sum_i\frac{n_i}{64\pi^2}m_i^4(h_1,h_2)
\left[\log\frac{m_i^2(h_1,h_2)}{Q^2}-C_i\right]
\ee
where $Q$ is the $\overline{\rm MS}$ 
renormalization scale, $C_i$=3/2 (5/6) for scalar bosons and fermions
(for gauge bosons), $m_i^2(h_1,h_2)$ is the field
dependent mass of the $i^{\rm th}$ particle in the background
$h_1$, $h_2$, and $n_i$ is the corresponding number of degrees
of freedom, which is taken negative for fermions. 

By minimizing the effective potential $V^{(0)}+V^{(1)}$ with
respect to ($h_1,h_2$), and imposing the minimum of the potential to be
at $(v_1,v_2)$, with $v=\sqrt{v_1^2+v_2^2}=174.1$ GeV, and 
$\tan\beta=v_2/v_1$ fixed, we can eliminate $m_1^2$ and $m_2^2$
in favour of the other parameters of the theory, 
while $m_3^2$ can be traded in favour of the one-loop corrected
squared mass $m_A^2$ of the CP-odd neutral Higgs boson~\cite{mariano2}.

Then we can write the finite temperature effective potential as
\be
V(h_1,h_2,T)=V^{(0)}(h_1,h_2)+V^{(1)}(h_1,h_2)+\Delta V(h_1,h_2,T)
\label{pottotal}
\ee
where the thermal correction $\Delta V$ contains the 
one-loop~\cite{mariano2}, $\Delta V^{(1)}(h_1,h_2,T)$,
plus the leading two-loop radiative corrections~\cite{JoseR,JRB,CQW2},
$\Delta V^{(2)}(h_1,h_2,T)$, 
of the daisy resummed theory. Since the two-loop corrections
have been proved to be very relevant for the strength of the
phase transition~\cite{JoseR}, and present estimates of bubble
parameters are based on one-loop thermal corrections, we will
often compare our numerical two-loop calculations of these parameters
with one-loop results.

The Euclidean action of configuration (\ref{background}) is given by:

\be
S_3(T)  =
4 \pi \bigint dr\; r^2
\left[ \left( \frac{d\, h_1}{d\, r} \right )^2  +
       \left( \frac{d\, h_2}{d\, r} \right )^2  + V(h_1, h_2, T)
\right]
\label{accion}
\ee
where $V(h_1, h_2, T)$ is the effective potential given 
in (\ref{pottotal}), shifted in such a way that 
\newpage\noindent
$V(0,0,T) = 0$.
The bubble is then the solution of the equations
%\newpage
%
\bea
\frac{d^2 h_1} {d r^2} + \frac{2}{r} \frac{d h_1}{d r} 
& = & \frac{1}{2} \frac{\partial V}{\partial h_1} \nonumber\\
\frac{d^2 h_2} {d r^2} + \frac{2}{r} \frac{d h_2}{d r}
& = & \frac{1}{2} \frac{\partial V}{\partial h_2} 
\label{burbuja}
\eea
supplied by the boundary conditions 

\be
\left.\frac{d h_i}{d r} \right|_{r=0} = 0;  \qquad 
\left.h_i\right|_{r=\infty} = 0\qquad (i=1,2)\ .
\label{frontera}
\ee
This solution only exists in the range $T_d<T<T_0$, where $T_d$ is
the temperature at which the minimum at $(h_1,h_2)$ is
degenerate with that at the origin, and $T_0$ is the temperature at
which the origin gets destabilized along some direction 
and becomes a saddle point.

Solving the previous differential equations is a difficult
task. There is no systematic procedure to find the solution. 
For example, the usual overshooting-undershooting method cannot be
implemented in the two (or more) field case. 
A possible technique is to define a functional, acting in some space, 
having its minimum at the solution of the previous equations. Notice
that the bubble solution is a saddle point of the Euclidean action,
{\em not} a minimum, so we cannot use directly $S_3$. One
possibility, explored in reference~\cite{kusenko}, is to add to $S_3$ some 
(nonlocal) terms that lift the falling directions. Then the
bubble is a minimum of this {\em improved} action. Here we will
follow a slightly  different approach. We will consider the following 
functional~\footnote{This kind of functionals has also been used
to obtain unstable solutions, such as sphalerons, on the 
lattice~\cite{marga}.}

\be
{\cal F}  = 
4 \pi \bigint dr\; r^2
\left[ 
   \left(
   \frac{d^2 h_1} {d r^2} + \frac{2}{r} \frac{d h_1}{d r}
-  \frac{1}{2} \frac{\partial V}{\partial h_1} 
   \right)^2 
+
   \left(
   \frac{d^2 h_2} {d r^2} + \frac{2}{r} \frac{d h_2}{d r}
-  \frac{1}{2} \frac{\partial V}{\partial h_2}
   \right)^2
\right]
\label{funcional}
\ee

By construction, the bubble solution is a minimum of the functional
${\cal F}$. In order to find numerically this solution, we have discretized 
the radial integral and spatial derivatives in the functional. The length
scale
used for the radial variable is $M_W^{-1}$. The bubble typically spreads over
a range of order 100 $M_W^{-1}$. Instead of minimizing simultaneously
$2N$ points describing the $h_1$ and $h_2$ profiles on the lattice,
we will use an iteration method. 
First, to have an estimate of the bubble size, we solve a
reduced problem. Assuming that the tunneling takes place along the
direction given by constant $\tan\beta(T)$, we can use the standard
overshooting-undershooting method to find numerically this 
configuration. 
This provides us with a good estimate of the size of the lattice,
where to solve the discretized problem, 
and also a first starting configuration for the minimization.
Now, we fix one of the Higgses, $h_1$ for example, to the previous estimate 
and allow the $N$ points
describing the other Higgs to change, minimizing the functional
given in (\ref{funcional}). We iterate this procedure, now fixing the
$h_2$ field. We have used $N=140$ for the discretization. 
In the range of the supersymmetric parameters we have explored,
we found a good convergence after two or three iterations in $h_1, h_2$.
Various tests derived from the fact that the obtained solutions are extremals
of the Euclidean action, as e.g. the virial theorem, have
been proved numerically to hold.
We have also verified stability of the profiles against 
increasing values of $N$.

%\vspace{1.5cm}
\nsect{Numerical results}

%\vspace{1cm}
The strength of the phase transition, after introduction of the
leading two-loop thermal corrections, has been studied in 
Refs.~\cite{JoseR}-\cite{CQW2}. The light
stop scenario was carefully analyzed in Ref.~\cite{CQW2}
where the effective potential along both the Higgs and $\st_R$ 
directions was considered, including the corresponding
two-loop corrections. There, the different possibilities
according to the cosmological evolution of the fields were
classified as regions of stability, metastability and
instability of the Higgs minimum, and regions where a two-step
phase transition can proceed. In this paper we concentrate
specifically in the stability region, since the metastability
and two-step regions would require considering the tunneling to (and
from) the color breaking minimum~\cite{MQS2}, which is outside the scope 
of the present paper. We will closely follow
the allowed parameter space found in Ref.~\cite{CQW2}. 

We choose $m_Q=1$ TeV in such a way that the supersymmetric
corrections to the $\rho$-parameter become small, giving hence a
good fit to the electroweak precision data coming from LEP and
SLC. On the other hand we will take as reference point
$\widetilde{A}_t\equiv A_t-\mu/\tan\beta=0$, 
$m_{\st}=150$ GeV and $\tan\beta=2.5$, $m_A=200$ GeV, 
providing $m_h=80$ GeV, which is consistent with present
experimental bounds on the MSSM lightest Higgs mass. This particular point
in the plane ($m_h,m_{\st}$) is acceptable from the point of
view of avoiding the wipe off of the baryon asymmetry after the
phase transition, as was shown in Figs.~1 and 2 of
Ref.~\cite{CQW2}. We have checked that the obtained results are
rather generic, as we will explicitly show by varying the
parameter $m_A$, so for other points in the allowed region of
the plane ($m_h,m_{\st}$) we obtain similar features.

In Fig.~1 we plot the euclidean action $S_3$ (\ref{accion}) for the
bounce solution (\ref{burbuja}) as a function of the temperature
for $\widetilde{A}_t=0$, $m_{\st}=150$ GeV, $\tan\beta=2.5$ and 
$m_A=200$ GeV. We can see from Fig.~1 that $S_3$ goes to
infinity at the degeneracy temperature $T_d\sim 96$ GeV, while
it goes to zero at the destabilization temperature $T_0\sim 93$
GeV. So all the phase transition happens in the 3 GeV interval
shown in Fig.~1. Using that
$S_3(T_c)/T_c\sim 
140-145$
one can easily compute the value of the temperature $T_c$~\footnote{The actual
temperature $T_c$ at which the
transition happens is readily computed by comparing the
probability of bubble nucleation per unit time and unit
volume, $\sim T^4\exp\{-S_3(T)/T\}$, with the expansion rate of the
universe at the corresponding temperature, 
$\tau^{-1}=\zeta^{-1}\; T^2/M_{\rm Pl}$, with $\zeta^{-1}=
4\pi\sqrt{\pi \left[g_{B}(T)+7/8\; 
g_F(T)\right]/45}$. By imposing the condition that the probability for a single
bubble to be nucleated within one horizon volume is $\cal O$(1)
one can compute the temperature $T_c$, i.e. from
$$\int_{T_c}^{\infty}\frac{dT}{T}\left(\frac{2\zeta M_{\rm
Pl}}{T}\right)^4 \exp\{-S_3(T)/T\}={\cal O}(1)\ .
$$}. For the case
considered in Fig.~1 we obtain $T_c\simeq 95.2$ GeV. The
corresponding profiles for the Higgs bubbles $h_1(r)$ and
$h_2(r)$ are plotted in Figs.~2 and 3 [where $\rho^2(r)\equiv
h_1^2(r)+ h_2^2(r)$ and $\tan\beta(r)=h_2(r)/h_1(r)$,
$\Delta\beta(r) =\beta(r)-\beta(T)$, $\beta(T)$ being
defined by $\beta(T)\equiv v_2(T)/v_1(T)$],  
for $T=T_c$ (solid lines), $T=T_c+0.4$ GeV (long-dashed lines) and
$T=T_c-0.4$ GeV (short-dashed lines). 

The variation of the bubble parameters with respect to $m_A$ is
displayed in Figs.~4, 5 and 6. In Fig.~4 the profile
$\rho(r)/v(T_c)$ is plotted for $m_A=100$, 200, 300 and 400 GeV,
thick-solid, long-dashed, short-dashed and thin-solid curves,
respectively. In all cases the bubbles have thick walls, the
wall thickness being $$L_\omega\sim (20-30)/T_c$$ as can be seen from
Fig.~4. We can also see that wall profiles are almost
indistinguisable for $m_A\simgt 200$ GeV. In Fig.~5 we plot the
parameter $\Delta\beta\equiv\Delta\beta(\infty)-\Delta\beta(0)$ 
as a function of $m_A$ for the whole
effective potential, including the two-loop thermal corrections
(solid line) and excluding them -i.e. including only the one-loop thermal 
corrections- (dashed line). We see there is an enhancement due
to the two-loop corrections which goes from $\sim$ 5 for
$m_A\sim 100$ GeV, to $\sim$ 2 for $m_A\sim 400$ GeV. Since the
total amount of produced baryon asymmetry is proportional to
$\Delta\beta$, the latter enhancement translates into a
baryogenesis enhancement, which was disregarded in previous
analyses~\cite{CQW2}. Furthermore, the enhancement produced in the 
strength of the phase transition, $v(T)/T$, 
by the two-loop thermal corrections~\cite{JoseR}-\cite{CQW2} 
also affects the amount of baryon asymmetry produced, since the
latter is proportional to~\cite{CQRVW} the integral
\be
I=\int_0^{\infty}
dr\;\frac{\rho^2(r)}{T^2}\;\frac{d\beta(r)}{dr} \ ,
\label{integral}
\ee
and therefore the two-loop enhancement is further strengthened. 
In Fig.~6 we
plot the integral (\ref{integral}) as a function of $m_A$, when
the two-loop thermal corrections are included (solid line) and
excluded (dashed line). In all cases the total enhancement is
greater than one order of magnitude.

\nsect{Squark bubbles}

In the previous analysis we have assumed that the bubble is 
built by the neutral field Higgses. This ansatz, used 
in the Euler-Lagrange equations, is obviously consistent.
However, if we want to interpret this solution as 
the one controlling the tunneling process, we have to 
analyze the variation of the Euclidean action under
generic quadratic perturbations including the rest of the fields. 
In particular, we have to make sure that there is just one negative mode:
the breathing mode.

The main candidate for such a new negative mode is provided
by squark fields, in particular stop fields. 
If we include these field in our analysis, then 
the finite temperature effective potential 
will be now a function of four fields,
$V_{eff} (h_1, h_2, \tilde{t}_L, \tilde{t}_R)$. In general, 
there will be directions involving the new fields 
$ \tilde{t}_L, \tilde{t}_R$ where the potential 
decreases~\footnote{This is certainly the case for some values 
of the parameters where new minima do appear.}. 
Since the kinetic term is always positive, 
negative modes imply necessarily negative values for the variation 
of the potential energy. In the quadratic approximation, this
variation is given by
\be
\label{curvature}
\delta^2 V_{\;\widetilde t} =4\pi\;
\bigint r^2 dr \left[
\begin{array}{cc}
 \widetilde{t}_L^*(r)&  \widetilde{t}_R^*(r)
\end{array}\right]
 {\cal M}^2_{\st}
\left[
\begin{array}{c}   
 \widetilde{t}_L(r) \\
 \widetilde{t}_R(r)
\end{array}\right]
\ee
where
\be
\label{m2sq}
{\cal M}^2_{\st}\;(h_1, h_2) =
\left[  \begin{array}{cc}
{\displaystyle m_Q^2 + \Pi_{\st_L}  + h_t^2 \, h_2^2(r)}
& A_t\;h_t \, h_2(r) - \mu\;h_t \,h_1(r)  
\\ & \\
  A_t\;h_t \, h_2(r) - \mu\;h_t \,h_1(r)  &
{\displaystyle m_U^2 + \Pi_{\st_R}  + h_t^2 \, h_2^2(r) }     
\end{array}
\right]
\ee
and $\Pi_{\st_{L,R}}$ stand for the leading $T-$dependent
self-energy contributions to the thermal masses. 
They are proportional to $T^2$~\cite{mariano2,CE}.
The eigenvalues of ${\cal M}^2_{\st}\;(h_1,h_2)$ give us the {\it masses} 
of the stop squarks in the Higgs background.
We have to evaluate this masses along
the path in the $(h_1, h_2)$ plane described by the bubble
when $r$ changes from zero to infinity.  
Since the variation of $\tan \beta$ 
along this path is less than $O(10^{-2})$, 
we will assume that $\widetilde{A}_t $ is
constant.  Notice that these masses must be positive
at the false and at the true vacuum because these are
a local and a global minimum, respectively. 
\newpage
These conditions
translate into:

\be
\begin{array}{rcl}
  m_Q^2 + \Pi_{\st_L}  & >&  0 \\ 
  m_U^2 + \Pi_{\st_R}  & >&  0 \\
 (m_Q^2 + \Pi_{\st_L}  + m_t^2)
 (m_U^2 + \Pi_{\st_R}  + m_t^2) & >&  \widetilde{A}_t^2\; m_t^2
\end{array}
\ee
On the other hand, a necessary condition 
for having negative masses along the path followed by the bubble
is:
\be
\widetilde{A}_t^2 > 
m_Q^2 + \Pi_{\st_L}  + m_U^2 + \Pi_{\st_R}   
\label{condicion}
\ee
However, this condition is forbidden by the requirement of not
wipping off, after the phase transition, any previously generated
baryon asymmetry \cite{CQW2}, and is never satisfied in our
choice of parameter space. We then conclude that the bubble
solution obtained in section 2 is not disturbed by
non trivial configurations for the fields
$\st_L$ and/or $\st_R$, and our conclusions in section 3 are
rather robust. Had we chosen a value of the mixing 
$\widetilde{A}_t^2$ satisfying condition (\ref{condicion}) we could have in the
bubble wall a field configuration with non-trivial values of $\st_R$ and/or
$\st_L$ explicitly violating baryon number.

\nsect{Conclusions}

In this paper we have computed the tunneling processes from the 
symmetric phase to the
true minimum in the first order phase transition of the Higgs fields in the
MSSM. We have obtained the corresponding Higgs profiles along the bubbles. We
paid particular attention  to the so-called light stop scenario, where the phase
transition is strong enough not to wipe off any previously generated baryon
asymmetry. Baryogenesis in the MSSM was previously proved to be controlled by
some bubble parameters, and in particular by the integral (\ref{integral}),
which were estimated by some energetic considerations based on the one-loop
effective potential. We have shown, using our numerical calculations based on
the two-loop effective potential, that there are important enhancement effects,
with respect to those estimates, that can be around one order of magnitude.
Finally we have proved that, for the considered cases, our solution is not
disturbed by any non-trivial configurations of the stop fields.

\vspace{1cm}
{\large\bf Acknowledgements}
 
\vspace{0.5cm}   
\noindent
We thank D. Oaknin for useful comments and for participating 
in the early stages of this work.

%%%%%%%%%%%%%%%%%%%%%%%%figure%%%%%%%%%%%%%%%%%%%%%%%%
\begin{figure}[b]
%\psdraft
\centerline{
\psfig{figure=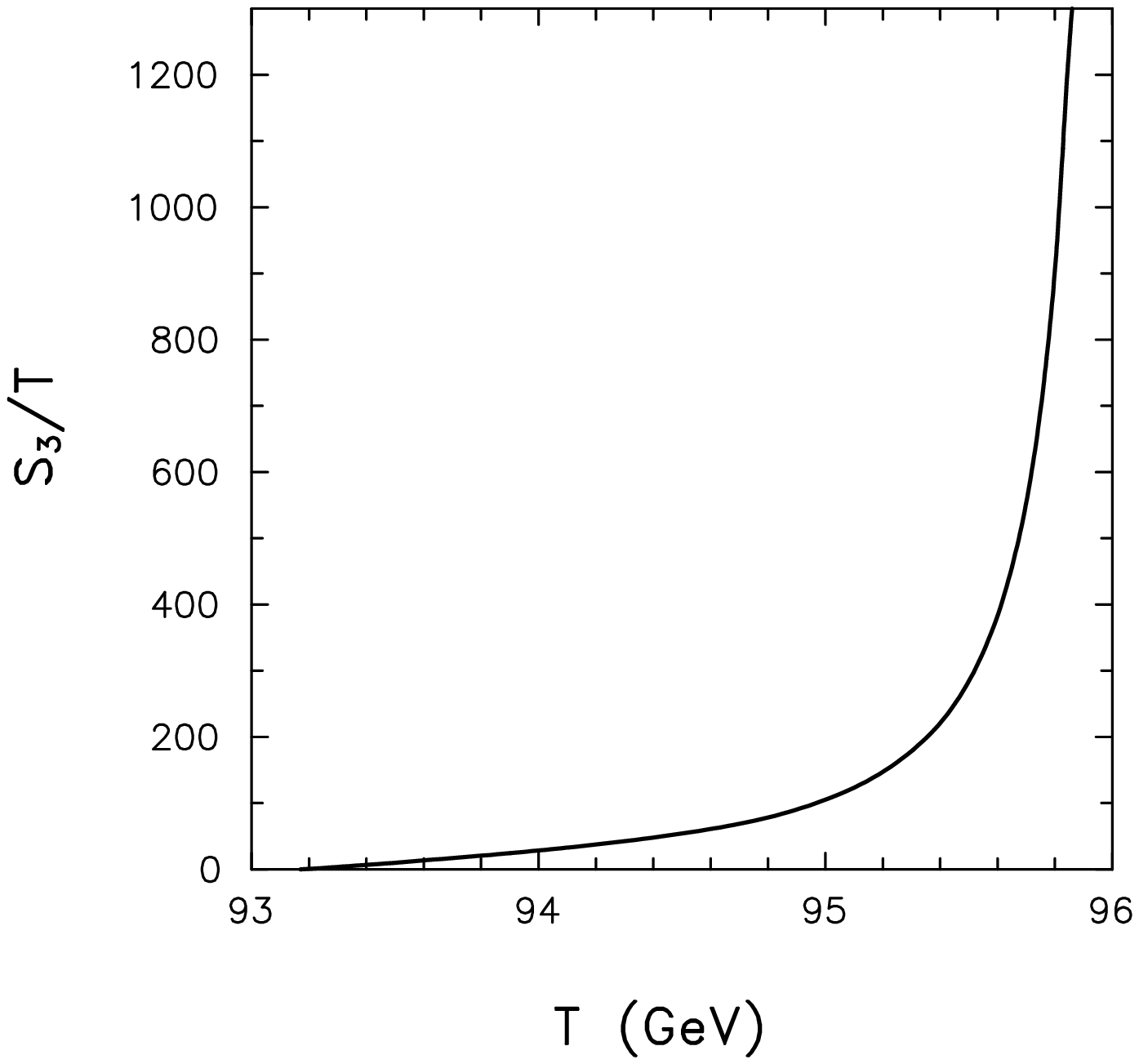,height=13cm,bbllx=5cm,bblly=3.5cm,bburx=19.cm,bbury=16.5cm}}
\caption{The Euclidean action as a function of the temperature for 
$m_Q=1$ TeV, $\tan\beta=2.5$, $\widetilde{A}_t=0$, $m_{\st}=150$ GeV, and
$m_A=200$ GeV.}
\label{f1}
\end{figure}
%%%%%%%%%%%%%%%%%%%%%%%%%%%%%%%%%%%figure%%%%%%%%%%%%%%%%%%%%%%%%
\begin{figure}[b]
%\psdraft   
\centerline{
\psfig{figure=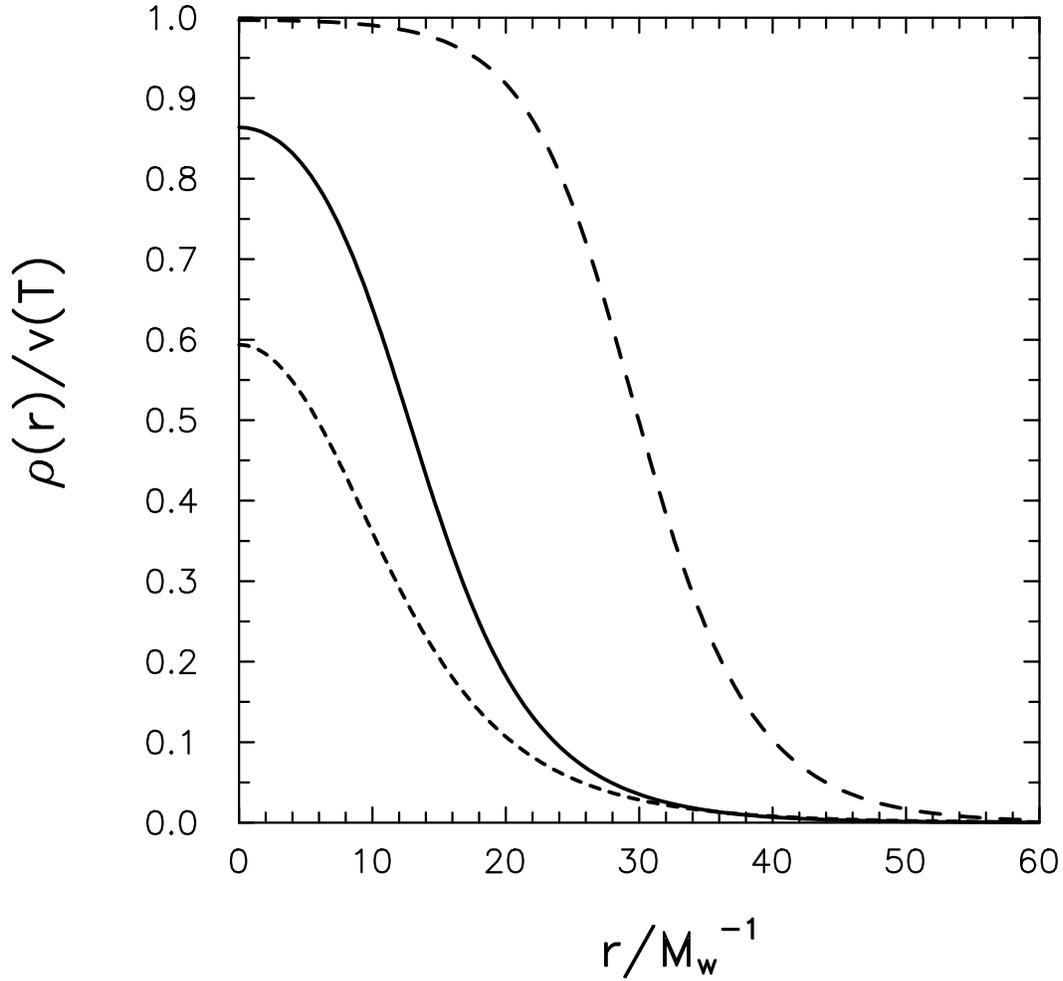,height=13cm,bbllx=5cm,bblly=3.5cm,bburx=19.cm,bbury=16.5cm}}
\caption{The Higgs profile $\rho(r)/v(T)$ for %T=T_c$ (solid curve),
$T=T_c+0.4$ GeV (long-dashed curve) and $T=T_c-0.4$ GeV (short-dashed curve),
and values of supersymmetric parameters as in Fig.~1.}
\label{f2}     
\end{figure}
%%%%%%%%%%%%%%%%%%%%%%%%%%%%%%%%%%%figure%%%%%%%%%%%%%%%%%%%%%%%%
\begin{figure}[b]
%\psdraft
\centerline{
\psfig{figure=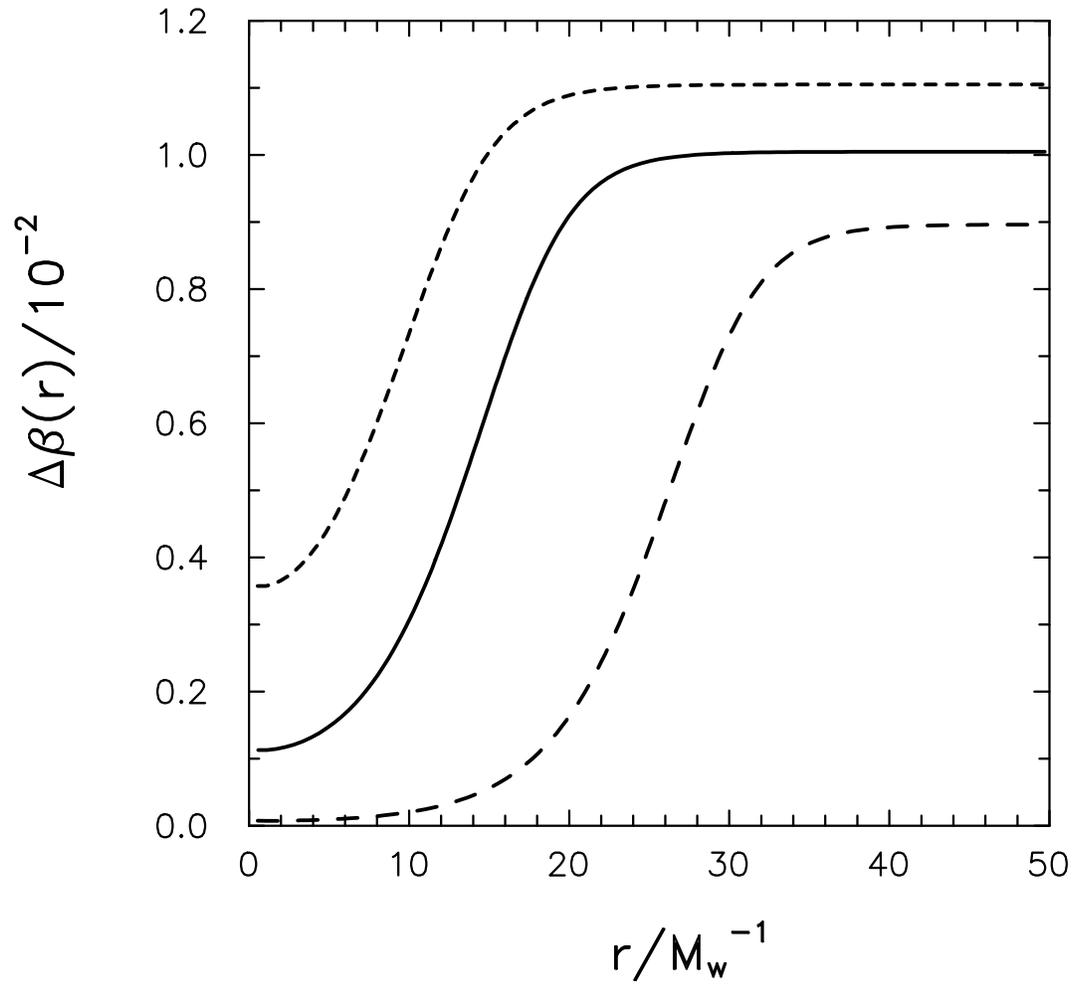,height=13cm,bbllx=5cm,bblly=3.5cm,bburx=19.cm,bbury=16.5cm}}
\caption{The same as in Fig.~2 but for the Higgs profile $\Delta\beta(r)$.}
\label{f3}
\end{figure}
%%%%%%%%%%%%%%%%%%%%%%%%%%%%%%%%%%%figure%%%%%%%%%%%%%%%%%%%%%%%%
\begin{figure}[b]
%\psdraft
\centerline{
\psfig{figure=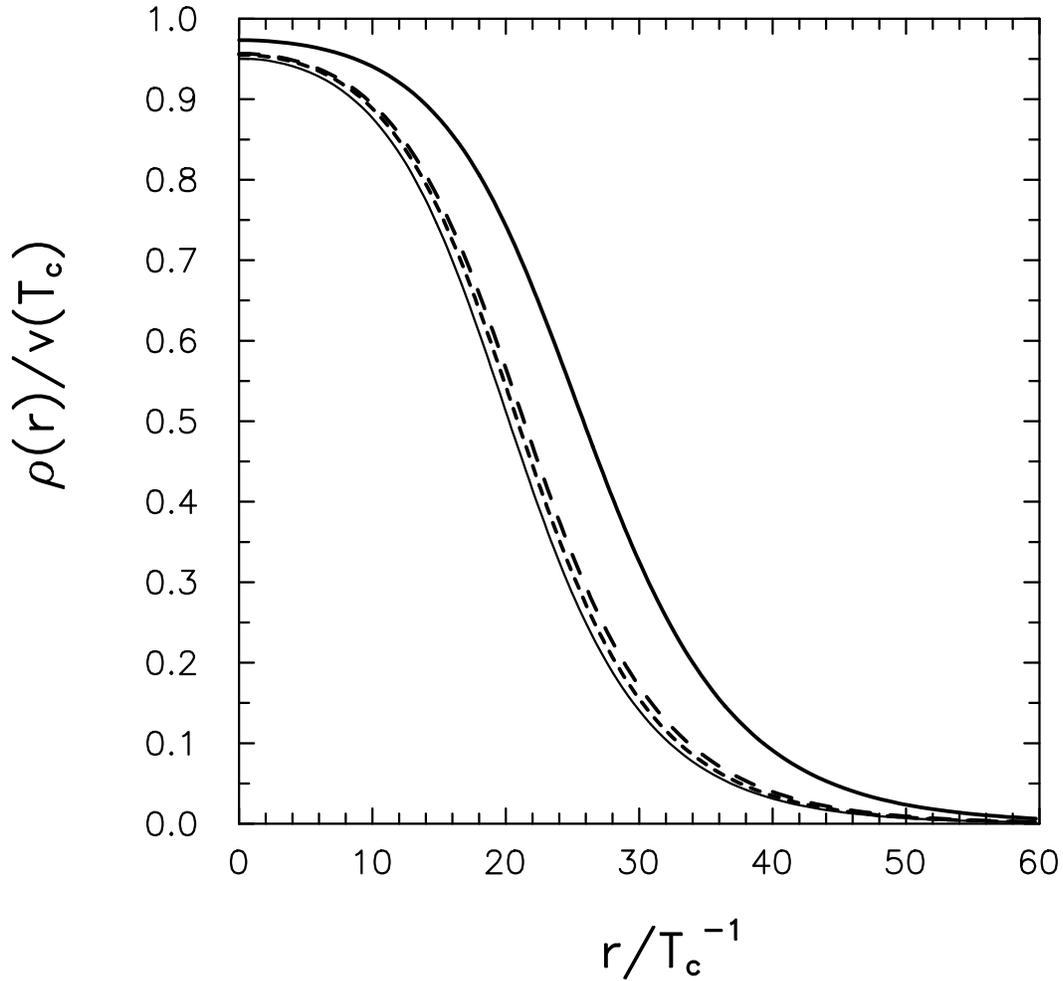,height=13cm,bbllx=5cm,bblly=3.5cm,bburx=19.cm,bbury=16.5cm}}
\caption{Higgs profile $\rho(r)/v(T_c)$ for $m_Q=1$ TeV, $\tan\beta=2.5$,
$\widetilde{A}_t=0$, $m_{\st}=150$ GeV, and $m_A=100$ GeV (thick-solid curve),
200 GeV (long-dashed curve), 300 GeV (short-dashed curve) and 400 GeV
(thin-solid curve).}
\label{f5}
\end{figure}
%%%%%%%%%%%%%%%%%%%%%%%%%%%%%%%%%%%figure%%%%%%%%%%%%%%%%%%%%%%%%%
\begin{figure}[b]
%\psdraft
\centerline{
\psfig{figure=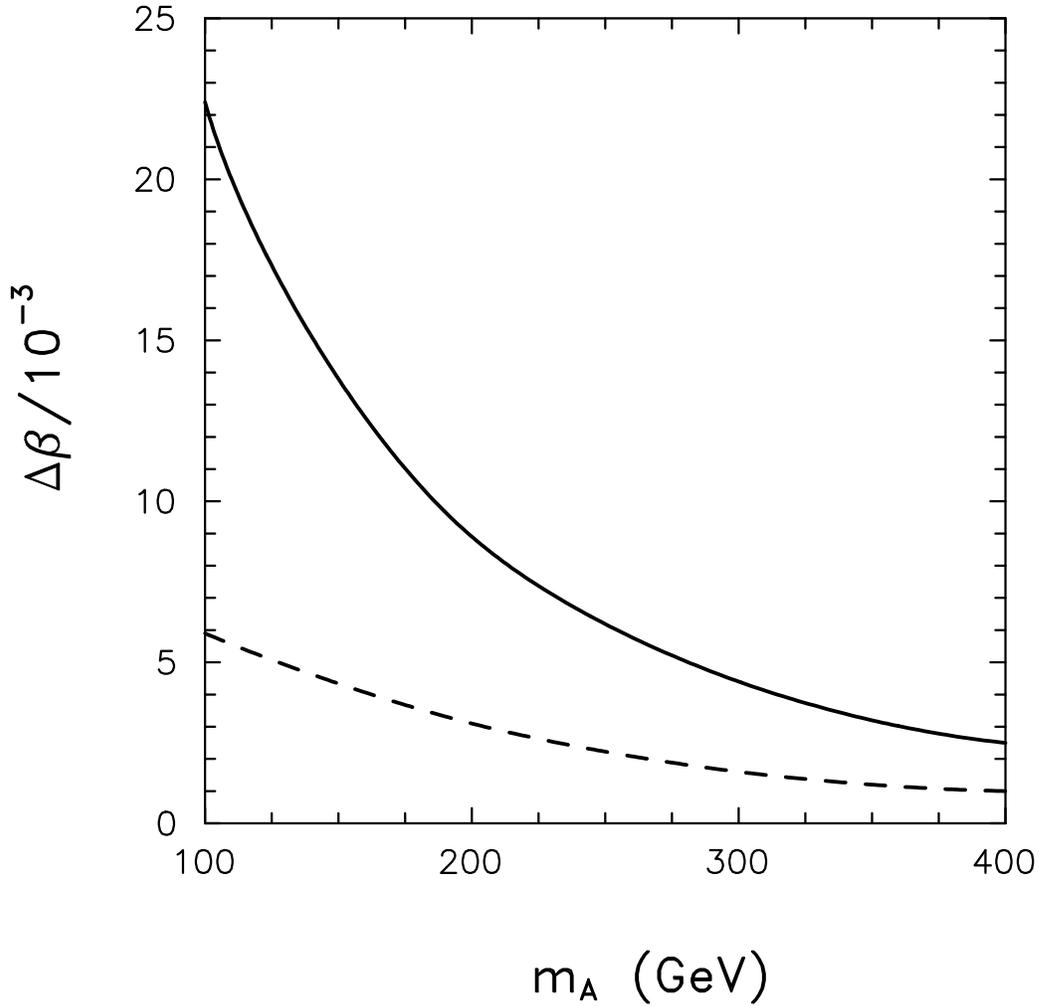,height=13cm,bbllx=5cm,bblly=3.5cm,bburx=19.cm,bbury=16.5cm}}
\caption{The parameter $\Delta\beta$ in the two-loop (solid curve) and one-loop
(dashed curve) approximations, for
$m_Q=1$ TeV, $\tan\beta=2.5$, $\widetilde{A}_t=0$ and $m_{\st}=150$ GeV, as a
function of $m_A$.}
\label{f6}
\end{figure}
%%%%%%%%%%%%%%%%%%%%%%%%%%%%%%%%%%%%%figure%%%%%%%%%%%%%%%%%%%%%%%%%
\begin{figure}[b]
%\psdraft
\centerline{
\psfig{figure=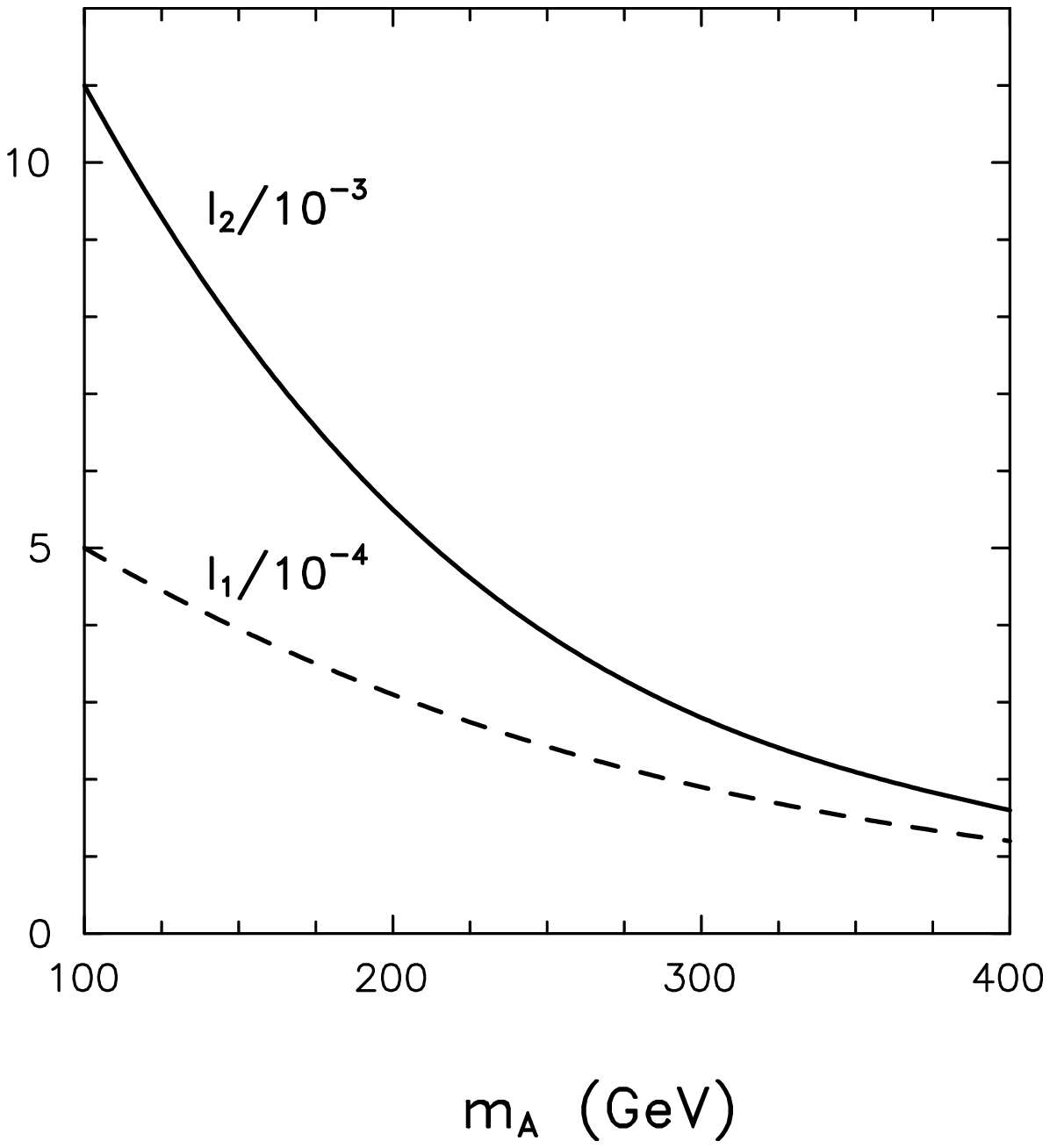,height=13cm,bbllx=5cm,bblly=3.5cm,bburx=19.cm,bbury=16.5cm}}
\caption{The same as in Fig.~5  but for the integral $I_{\ell}$
(3.1), $\ell$=1-loop (dashed curve), $\ell$=2-loop (solid curve).}
\label{f7}
\end{figure}
%%%%%%%%%%%%%%%%%%%%%%%%%%%%%%%%%%%%%%%%%%%%%%%%%%%%%%%%%%%%%%%%%

\newpage
\begin{thebibliography}{99}
%
\bibitem{baryogenesis} A.D.~Sakharov, \JETPL{91B}{67}{24}
%
\bibitem{first} M. Shaposhnikov, 
{\it JETP Lett.} 44 (1986) 465; \NPB{287}{87}{757} and
{\bf B299} (1988) 797. P.~Arnold and L.~McLerran, \PRD{36}{87}{581};
and {\bf D37} (1988) 1020; S.Yu~Khlebnikov and M.E.~Shaposhnikov,
\NPB{308}{88}{885};
F.R. Klinkhamer and N.S. Manton, \PRD{30}{84}{2212};
B. Kastening, R.D. Peccei and X. Zhang, \PLB{266}{91}{413};
L.~Carson, Xu~Li, L.~McLerran and R.-T.~Wang, \PRD{42}{90}{2127};
M.~Dine, P.~Huet and R.~Singleton Jr., \NPB{375}{92}{625}
%
\bibitem{reviews} For recent reviews, see:
A.G. Cohen, D.B. Kaplan and A.E. Nelson,
{\it Annu. Rev. Nucl. Part. Sci.} {\bf 43} (1993) 27;
M. Quir{\'o}s, \HPA{67}{94}{451}; V.A.~Rubakov and M.E.~Shaposhnikov,
{\it Phys. Usp.} {\bf 39} (1996) 461-502 [hep-ph/9603208]
%
\bibitem{AndH} G.W. Anderson and L.J. Hall, \PRD{45}{92}{2685}.
%
\bibitem{improvement} M.E. Carrington, \PRD{45}{92}{2933};
M. Dine, R.G. Leigh, P. Huet, A. Linde and D. Linde, \PLB{283}{92}{319};
\PRD{46}{92}{550}; P. Arnold, \PRD{46}{92}{2628};
J.R. Espinosa, M. Quir{\'o}s and F. Zwirner, \PLB{314}{93}{206};
W. Buchm{\"u}ller, Z. Fodor, T. Helbig and D. Walliser, \AP{234}{94}{260}
%
\bibitem{twoloop} J.~Bagnasco and M.~Dine, \PLB{303}{93}{308};
P. Arnold and O. Espinosa, \PRD{47}{93}{3546}; Z. Fodor and A. Hebecker,
\NPB{432}{94}{127}
%
\bibitem{nonpert} K. Kajantie, K.~Rummukainen
and M.E.~Shaposhnikov, \NPB{407}{93}{356};
 Z. Fodor, J. Hein, K. Jansen, A. Jaster and
I. Montvay, \NPB{439}{95}{147};
K.~Kajantie, M.~Laine, K.~Rummukainen and
M.E.~Shaposhnikov, \NPB{466}{96}{189};
K.~Jansen, {\it Nucl. Phys. (Proc. Supl.)} {\bf B47} (1996) 196;
K. Rummukainen, {\it Nucl. Phys. (Proc. Supl.)}
{\bf B53} (1997) 30
%
\bibitem{CPSM} G.R.~Farrar and M.E.~Shaposhnikov, \PRL{70}{93}{2833} and
({\bf E}): {\bf 71} (1993) 210; M.B.~Gavela. P.~Hern{\'a}ndez, J.~Orloff,
O.~P{\`e}ne and C.~Quimbay, \MPLA{9}{94}{795};
\NPB{430}{94}{382}; P.~Huet and E.~Sather, \PRD{51}{95}{379}
%
\bibitem{CPMSSM} M.~Dine, P.~Huet, R.~Singleton Jr. and L.~Susskind,
\PLB{257}{91}{351}; A.~Cohen and A.E.~Nelson, \PLB{297}{92}{111};
P. Huet and A.E. Nelson, \PLB{355}{95}{229};
\PRD{53}{96}
%
\bibitem{CQRVW} M. Carena, M. Quiros, A. Riotto, I. Vilja
and C.E.M. Wagner, \NPB{503}{97}{387}
%
\bibitem{CJK} J. Cline, M. Joyce and K. Kainulainen,
preprint MCGILL-97-26, [hep-ph/9708393]
%
\bibitem{Iiro2} T. Multamaki, I. Vilja, \PLB{411}{97}{301}
%
\bibitem{Toni2} A. Riotto, preprint OUTP-97-43-P,
[hep-ph/9709286]
%
\bibitem{Worah} M. Worah, \PRD{56}{97}{2010}
%
\bibitem{ReviewCW} M. Carena and C.E.M. Wagner,
preprint FERMILAB-PUB-97-095-T, [hep-ph/9704347],
to appear in 'Perspectives on Higgs Physics II', ed. G. Kane,
World Scientific, Singapore, 1997
%
\bibitem{last} H.~Davoudiasl, K.~Rajagopal and E.~Westphal, preprint
CALT-68-2127 [hep-ph/9707540] 
%
\bibitem{early} G.F. Giudice, \PRD{45}{92}{3177};
S. Myint, \PLB{287}{92}{325}
%
\bibitem{mariano1} J.R. Espinosa, M. Quir{\'o}s and F. Zwirner,
\PLB{307}{93}{106}
%
\bibitem{mariano2} A. Brignole, J.R. Espinosa, M. Quir{\'o}s and F. Zwirner,
\PLB{324}{94}{181}
%
\bibitem{CQW} M. Carena, M. Quir{\'o}s and C.E.M. Wagner,
\PLB{380}{96}{81}
%
\bibitem{Delepine} D. Delepine, J.M. G\'erard, R. Gonz\'alez Felipe
and J. Weyers, \PLB{386}{96}{183}
%
\bibitem{MOQ} J.M. Moreno, D.H. Oaknin and M. Quir{\'o}s, \NPB{483}{97}{267};
\PLB{395}{97}{234}
%
\bibitem{CK} J. Cline and K. Kainulainen, \NPB{482}{96}{73};
preprint MCGILL-97-7, [hep-ph/9705201]
%
\bibitem{FL} M. Laine, \NPB{481}{96}{43};
M. Losada, \PRD{56}{97}{2893}; preprint [hep-ph/9612337];
G. Farrar and M. Losada, \PLB{406}{97}{60}
%
\bibitem{JoseR} J.R. Espinosa, \NPB{475}{96}{273}
%
\bibitem{JRB} B. de Carlos and J.R. Espinosa, \NPB{503}{97}{24}
%
\bibitem{Schmidt} D. Bodeker, P. John, M. Laine and M.G. Schmidt,
\NPB{497}{97}{387}
%
\bibitem{CQW2} M. Carena, M. Quiros and C.E.M. Wagner, preprint
FNAL-Pub-97/327-T, CERN-TH/97-190, IEM-FT-165/97 [hep-ph/9710401]
%
\bibitem{Brihaye93}Y. Brihaye and J. Kunz \PRD{48}{93}{3884}.
%
\bibitem{Col78}
S. Coleman, V. Glaser and A. Martin, 
{\it Commun. Math. Phys. } {\bf 58} (1978) 211
%
\bibitem{kusenko} A. Kusenko, \PLB{358}{95}{51}
%
%
\bibitem{marga}
M.~Garc{\'\i}a-P{\'e}rez and P.~van Baal, \NPB{468}{96}{277}
%
\bibitem{MQS2} J.M. Moreno, M. Quir{\'o}s and M. Seco, in
preparation 
%
\bibitem{CE} D. Comelli and J.R. Espinosa, \PRD{55}{97}{6253}
\end{thebibliography}
\end{document}